\documentclass[12pt]{article}
\usepackage{amsmath}
\usepackage{bm}
\usepackage{color}
\usepackage{braket}
\usepackage{caption}
\usepackage{subcaption}
\usepackage{float}
\usepackage{graphicx}
\usepackage{amssymb}
\usepackage{mathrsfs}
\usepackage{amsmath}
\usepackage{tikz}
\usetikzlibrary{quantikz2}
\begin{document}
\begin{center}
\begin{large}
{\bf Stuctural properties of multi-partite graphs and quantum correlators}
\end{large}
\end{center}

\centerline {Kh. P. Gnatenko \footnote{E-Mail address: khrystyna.gnatenko@gmail.com}}
\medskip
\centerline {\small \it Ivan Franko National University of Lviv,}
\centerline {\small \it Professor Ivan Vakarchuk Department for Theoretical Physics,}
\centerline {\small \it 12 Drahomanov St., Lviv, 79005, Ukraine}

\abstract{We propose a method for constructing quantum graph states that can be represented by vertex- and edge-weighted multipartite graphs. Quantum correlators are calculated for the general case of quantum graph states corresponding to multipartite graphs of arbitrary structure. We find the relation between quantum correlators and the neighborhood structure of the selected vertices. In  particular two-point spin correlators are related to the number of four-cycles in the graph and the number of complete bipartite subgraphs $K_{2,m}$ formed by two selected vertices and $m$ vertices from a different partition of the graph. On the basis of the obtained results, a method for quantifying these structural properties using quantum computing is proposed. As an example, the number of four-cycles is quantified using quantum computing for a quantum graph state representing the $K_{2,3}$ graph.

}

\section{Introduction}

Quantum states that can be represented by graphs are known in the 
literature as quantum graph states. Studies of the properties of such 
states have recently attracted much attention 
\cite{Markham,Wang,Ghosh25, Mooney, Zan25, Bell,Mazurek,Shettell,Hein,
Guhne,Vesperini1,Vesperini2}.
Quantum graph states constructed with the help of controlled $Z$ gates 
have been well studied (see, for example, 
\cite{Wang,Mooney,Alba,Mezher,Akhound,Haddadi,Cabello}). 
The states read 
$\ket{\psi_G} = \prod_{(i,j) \in E} CZ_{ij}\ket{+}^{\otimes |V|}$, 
where $CZ_{ij}$ is the controlled-Z gate acting on qubits $q[i]$ 
and $q[j]$. These states are multi-qubit entangled states that can 
be related to unweighted graphs $G(V,E)$. In this case, each qubit 
is represented by a vertex of the graph, and the action of a 
two-qubit controlled-Z gate corresponds to an edge of the graph. 
In papers \cite{Susulovska}, a generalization of these 
states was studied. Namely, quantum graph states constructed by 
applying controlled-phase gates $CP_{ij}(\phi)$ to a separable 
multi-qubit state were considered, and the entanglement of the 
states was quantified both analytically and with the help of 
quantum computing.

Due to multipartite entanglement, efficient generation and other properties  quantum graph states are widely used in 
quantum computing, namely, in quantum error correction algorithms 
\cite{Bell,Mazurek,Hillmann}, quantum machine learning 
\cite{Gao,Zoufal}, quantum cryptography 
\cite{Markham, Sha}. 

In this paper, we consider a more general notion of a quantum graph 
state. In particular, our approach is not limited to the construction of quantum graph states using $CZ$ or $CP(\phi)$ gates. We consider a quantum 
graph state to be a state prepared in such a way that a unique 
graph representation of the state exists. In our previous paper, 
we studied quantum graph states corresponding to the evolutionary states of spin 
systems  with Ising model that can be represented by 
weighted and directed graphs (see, \cite{Gnatenko} and references therein). Bipartite quantum 
graph states constructed by the action of controlled-NOT gates 
were also studied. The entanglement of the states was quantified, 
and the relation of quantum correlators to the number of vertices 
of odd and even degrees was examined \cite{Gnatenko26}. In \cite{{Gnatenko26IEEE}}  tripartite 
quantum graph states   constructed by the action of $RXY$, $RYZ$, 
and $RXZ$ gates on a separable quantum state were examined. The
relation between the properties of the states and those of the 
corresponding graphs was found.

In this paper, we propose a way to construct quantum states that 
correspond to weighted $k$-partite graphs. An $k$-partite graph 
is a graph whose vertices can be divided into $k$ subsets such 
that there are no edges between vertices belonging to the same 
subset. Such graphs are widely used in problems of resource 
allocation, scheduling, and modeling of complex systems.
 It is found that quantum correlations in the states
depend on 
distribution of the neighbors of the selected vertices among 
different partitions, the vertex degrees, and the number of 
four-cycles. As a result, a methods for quantifying the number of 
four-cycles in multipartite graphs is developed. The obtained results open a possibility to determine the number of complete bipartite subgraphs $K_{2,m}$ formed by two selected vertices and $m$ vertices from another partition of the graph with quantum computing. 

The paper is organized as follows. In Section 2, a method for 
constructing quantum states that can be represented by multipartite 
graphs is presented. In Section 3, quantum correlators in quantum 
graph states are calculated for an arbitrary structure of the 
graph. We establish the relation between quantum correlators and 
the structural properties of the graphs. In Section 4, the number 
of four-cycles is quantified using quantum computing in particular case of $K_{2,3}$ graph. Conclusions 
are presented in Section 5.

\section{Quantum states representing multipartite weighted graphs}

Let us consider multi-qubit quantum states that can be represented by  edge- and vertex-weighted multipartite graphs. The states read
\begin{eqnarray}
\ket{\psi_G}=\prod_{(i,j) \in E} RXX_{ij}(\phi_{ij}) \prod_{k \in V} (\cos \frac{\theta_k}{2} \ket{0}_k + e^{\mathrm{i}\alpha_k} \sin \frac{\theta_k}{2} \ket{1}_k). \label{eq:2:1}
\end{eqnarray}
Here operators $RXX_{ij}(\phi_{ij})$ are defined as
$RXX_{ij}(\phi_{ij})=\exp(-\mathrm{i}\phi_{ij}\sigma^x_i\sigma^x_j/2)$, where $\sigma^x_i$, $\sigma^x_j$ are Pauli matrices corresponding to qubits $q[i]$ and $q[j]$.
The state can be considered as multipartite quantum graph state representing graph $G=(V,E,P)$ with partitions 
\begin{eqnarray}
P=\{V_1,\ldots,V_p\}, \ \ V=\bigcup_{i=1}^{p} V_i.
\end{eqnarray}

Qubits represent the vertices of the graph, while the action of the two-qubit gate $RXX_{ij}(\phi_{ij})$ represents an edge between vertices $i$ and $j$. The weights of the edges are related to the parameters $\phi_{ij}$. In the more general case of directed graphs, the parameters $\phi_{ij}$ could also identify the direction of the corresponding edges. The arc $(i,j)$ can be represented by $\phi_{ij}$ and the arc $(j,i)$ by $\phi_{ji}$. The weight of vertex $i$ is related to the parameter $\theta_i$ in the initial state of qubit $q[i]$. To differentiate between partitions, the parameters $\alpha$ in the initial state are used. The parameter $\alpha_r$ is common to all vertices belonging to the subset $V_r$. Note that, up to a global phase, the initial state can be prepared by the action of the gates
$RZ(\alpha_j)=\exp\left(-\mathrm{i}{\alpha{j}}\sigma^z_j/2\right)$, $RY(\theta_j)=\exp\left(-\mathrm{i}{\theta{j}}\sigma^y_j/2\right)$ with  $\sigma^{y}_j$, $\sigma^{z}_j$ being the Pauli matrices corresponding to qubit $q[j]$.  We have
\begin{eqnarray}
RZ(\alpha_j)RY(\theta_j)\ket 0_j=e^{-\mathrm{i}\frac{\alpha_j}{2}}(\cos \frac{\theta_j}{2} \ket{0}_j + e^{\mathrm{i}\alpha_j} \sin \frac{\theta_j}{2} \ket{1}_j)
\end{eqnarray}
 At the end of this section, it is worth noting that the quantum graph states are the evolved quantum states of spin systems described by the Ising model.

In the next section, we calculate quantum correlators in the general case of multipartite graphs of arbitrary structure and study their relation to the structural properties of the corresponding graphs. 

\section{Quantum correlators and structural properties of multi-partite graphs}

Let us quantify quantum correlators in quantum state corresponding to general structure of multipartite graph and study their relations with the properties of corresponding graphs.
Let us first consider
\begin{eqnarray}
\bra{\psi_G}\prod_{i=1}^{k}\sigma_{i}^{y} \ket{\psi_G}=  \prod_{d \in V} \left(\cos \frac{\theta_d}{2} \bra{0}_d + e^{-\mathrm{i}\alpha_d} \sin \frac{\theta_d}{2} \bra{1}_d\right) \prod_{(a,b) \in E} RXX^+_{ab}(\phi_{ab}) \times \nonumber\\ \times \prod_{i=1}^{k}\sigma_{i}^{y} \prod_{(l,r) \in E} RXX_{lr}(\phi_{lr}) \prod_{v \in V} \left(\cos \frac{\theta_v}{2} \ket{0}_v + e^{\mathrm{i}\alpha_v} \sin \frac{\theta_v}{2} \ket{1}_v\right).
\end{eqnarray}
For convenience we introduce the following notations.
Let set $A$,  $A=\{i_1,i_2,\ldots,i_k\}\subseteq V$
represent vertices in graph $G(V,E,P)$ corresponding to the $k$
qubits involved in the correlator,
$A_i$ denotes the subset of vertices 
belonging to $V_i$, $A_i=A\cap V_i$.
Let also use notation $A_{ext}$ for the set of all vertices outside of those belonging to $A$ that are  connected with an edge to at least one vertex in $A$.
Namely, this external vertex
boundary of $A$ is defined as
$A_{ext}=
\{j\in V\setminus A:\ N(j)\cap A\neq\varnothing\}$,  $N(j)$ are
neighbors of vertex $j$. So, taking into account that
$\sigma_l^yRXX_{lr}(\phi_{lr})=RXX_lr^+(\phi_{lr})\sigma_l^y$ because ${\sigma^y_l, \sigma^x_l}=0$ we find 
\begin{eqnarray}
\prod_{(a,b) \in E} RXX^+_{ab}(\phi_{ab}) \prod_{i=1}^{k}\sigma_{i}^{y} \prod_{(l,r) \in E} RXX_{lr}(\phi_{lr})=
\nonumber\\=
\prod_{j\in A_{\mathrm{ext}}}
\prod_{m\in A\cap N(j)}
RXX_{mj}^{\dagger}(2\phi_{mj})
\prod_{i=1}^{k}\sigma_{i}^{y}.
\end{eqnarray} 
So, finally, for the quantum correlator we obtain
\begin{eqnarray}
\left\langle
\prod_{i=1}^{k}\sigma_{i}^{y}
\right\rangle
=\nonumber\\
\frac{1}{2^k}
\sum_{\beta_{1}=\pm1}
\sum_{\beta_{2}=\pm1}
\cdots
\sum_{\beta_{k}=\pm1}
\prod_{r=1}^{p}
\prod_{i\in A_r}
\left(
\sin\theta_{i}\sin\alpha_r
+\mathrm{i}\beta_{i}\cos\theta_{i}
\right)\times
\nonumber\\
\times
\prod_{s=1}^{p}
\prod_{j\in A_{ext}\cap V_s}
\left(
\cos \Phi_{sj}
+\mathrm{i}\sin\theta_j\cos\alpha_s
\sin
\Phi_{sj}
\right),
\label{eq:multipartite_y_correlator}
\end{eqnarray}
where 
\begin{eqnarray}
\Phi_{rj}=\sum_{\substack{s=1}}^{p}
\sum_{\substack{i\in A_s\cap N(j)}}
\beta_{i}\phi_{ij}(1-\delta_{sr}).
\end{eqnarray}

Similarly, for $\left\langle
\prod_{i=1}^{k}\sigma_{i}^{z}
\right\rangle$ we find

\begin{eqnarray}
\left\langle
\prod_{i=1}^{k}\sigma_{i}^{z}
\right\rangle
=\nonumber\\=\prod_{d \in V} \left(\cos \frac{\theta_d}{2} \bra{0}_d + e^{-i\alpha_d} \sin \frac{\theta_d}{2} \bra{1}_d\right) \prod_{j\in A_{\mathrm{ext}}}
\prod_{m\in A\cap N(j)}
RXX_{mj}^{\dagger}(2\phi_{mj})
\prod_{i=1}^{k}\sigma_{i}^{z} \times \nonumber \\ \times \prod_{v \in V} \left(\cos \frac{\theta_v}{2} \ket{0}_v + e^{i\alpha_v} \sin \frac{\theta_v}{2} \ket{1}_v\right)=\nonumber\\=
\frac{1}{2^k}
\sum_{\beta_{1}=\pm1}
\sum_{\beta_{2}=\pm1}
\cdots
\sum_{\beta_{k}=\pm1}
\prod_{r=1}^{p}
\prod_{i\in A_r}
\left(
\cos\theta_{i}
-\mathrm{i}\beta_{i}\sin\theta_{i}\sin\alpha_r
\right)\times
\nonumber\\
\times
\prod_{s=1}^{p}
\prod_{j\in A_{\mathrm{ext}}\cap V_s}
\left(
\cos\Phi_{sj}
+\mathrm{i}\sin\theta_j\cos\alpha_r
\sin\Phi_{sj}
\right).
\label{eq:multipartite_z_correlator}
\end{eqnarray}

And becasuse $[\sigma_i^x, RXX_{jk}(\theta_{jk})]=0$, for $\left\langle
\prod_{i=1}^{k}\sigma_i^x
\right\rangle$ we can write
\begin{eqnarray}
\left\langle
\prod_{i=1}^{k}\sigma_i^x
\right\rangle
=
\prod_{r=1}^{p}
\prod_{i\in A_r}
\sin\theta_i\cos\alpha_r.
\label{cxxx}
\end{eqnarray}
It is worth noting that
$\langle\prod^k_{i=1}\sigma_i^y\rangle$ and
$\langle \prod^k_{i=1} \sigma_i^z\rangle$  are determined
by the wights of the vertices belonging to
the selected set $A$,  and weights of verticies forming the
external neighbors $A_{\mathrm{ext}}$, and the
weights of the edges connecting them.

In general, for the quantum correlator
$\left\langle
\prod_{i\in X}\sigma_i^x
\prod_{j\in Y}\sigma_j^y
\prod_{l\in Z}\sigma_l^z
\right\rangle$,
let $X,Y,Z\subseteq V$ be pairwise disjoint subsets
of vertices corresponding to the qubits on which
the Pauli operators $\sigma^x$, $\sigma^y$, and
$\sigma^z$ act, respectively. We obtain
\begin{eqnarray}
\left\langle
\prod_{i\in X}\sigma_i^x
\prod_{j\in Y}\sigma_j^y
\prod_{l\in Z}\sigma_l^z
\right\rangle
=\nonumber\\
\frac{1}{2^{|Y|+|Z|}}
\prod_{i\in B}\sum_{\beta_i=\pm1}
\prod_{r=1}^{p}
\prod_{j\in Y\cap V_r}
\left(
\sin\theta_j\sin\alpha_r
+\mathrm{i}\beta_j\cos\theta_j
\right)
\nonumber\\
\times
\prod_{s=1}^{p}
\prod_{l\in Z\cap V_s}
\left(
\cos\theta_l
-\mathrm{i}\beta_l\sin\theta_l\sin\alpha_s
\right)
\prod_{t=1}^{p}
\prod_{m\in X\cap V_t}
\left(
\sin\theta_m\cos\alpha_t\cos\tilde{\Phi}_m
+\mathrm{i}\sin\tilde{\Phi}_m
\right)
\nonumber\\
\times
\prod_{v=1}^{p}
\prod_{n\in(B_{\mathrm{ext}}\setminus X)\cap V_v}
\left(
\cos\tilde{\Phi}_n
+\mathrm{i}\sin\theta_n\cos\alpha_v
\sin\tilde{\Phi}_n
\right).
\label{eq:general_pauli_correlator}
\end{eqnarray}

Here we use the following notations. Because, $\sigma_i^y$ and  $\sigma_i^z$ 
do not commute with $RXX_{ij}(\theta_{ij})$ we consider notation $B$ for $B=Y\cup Z$.
Also, $B_{\mathrm{ext}}$ reads
$B_{\mathrm{ext}}=
\{j\in V\setminus B:
N(j)\cap B\neq\varnothing\},$
and 
\begin{eqnarray}
\tilde{\Phi}_{rj}=
\sum_{s=1}^{p}
\sum_{i\in B\cap V_s\cap N(j)}
\beta_i\phi_{ij}(1-\delta_{sr}).
\end{eqnarray}

Let us study particular case of $\theta_i=\theta$ and $\phi_{ij}=\phi$.
It is interesting to note that in this case the correlator $\left\langle\sigma_a^z\sigma_b^z\right\rangle$, $a,b\in V_u$ reads

\begin{eqnarray}
\left\langle\sigma_a^z\sigma_b^z\right\rangle
=\frac{1}{2}\Re(
(\cos\theta-i\sin\theta\sin\alpha_u)^2 
\nonumber\\
\times\prod_{\substack{r=1\\r\neq u}}^{p}
\left(\cos\phi+i\sin\theta\cos\alpha_r
\sin\phi\right)^{
|(N(a)\setminus N(b))\cap V_r|
+
|(N(b)\setminus N(a))\cap V_r|
}
\nonumber\\
\times\prod_{\substack{r=1\\r\neq u}}^{p}
\left(\cos2\phi+i\sin\theta\cos\alpha_r
\sin2\phi\right)^{
|N(a)\cap N(b)\cap V_r|}
\nonumber\\
+
(\cos\theta-i\sin\theta\sin\alpha_u)
(\cos\theta+i\sin\theta\sin\alpha_u)
\nonumber\\
\times\prod_{\substack{r=1\\r\neq u}}^{p}
\left(\cos\phi+i\sin\theta\cos\alpha_r
\sin\phi\right)^{
|(N(a)\setminus N(b))\cap V_r|}
\nonumber\\
\times\prod_{\substack{r=1\\r\neq u}}^{p}
\left(\cos\phi-i\sin\theta\cos\alpha_r
\sin\phi\right)^{
|(N(b)\setminus N(a))\cap V_r|}).
\end{eqnarray}

So, we have dependence of the correlator on the 
 distribution of
neighbors of the selected vertices among the different
partitions of the multipartite graph.
Number $\left|(N(a)\setminus N(b))\cap V_r\right|$
corresponds to the number of vertices in partition $V_r$
that are adjacent to vertex $a$ but not to vertex $b$, $
\left|(N(b)\setminus N(a))\cap V_r\right|$ is the
number of vertices in $V_r$ that are
adjacent to vertex $b$ but not to vertex $a$, 
$\left|N(a)\cap N(b)\cap V_r\right|$ is the number of common neighbors of the vertices $a$ and $b$ that belong to partition $V_r$.
And 
$\left|(N(a)\setminus N(b))\cap V_r\right|
+
\left|(N(b)\setminus N(a))\cap V_r\right|$
is  the number of vertices in $V_r$
connected to exactly one of the selected vertices.

It is important to note that each pair of distinct common neighbors of
vertices $a$ and $b$ forms a cycle of length
four. So, knowing the value 
$|N(a)\cap N(b)\cap V_r|$
one can calculate the number of
four-cycles $C_4$ with vertices $a$ and $b$ as their opposite
vertices, and vertices belonging to the subset $V_r$, $r\neq u$
 This value reads
\begin{equation}
N^{(ur)}_{C_4}(a,b)
=
\binom{|N(a)\cap N(b)\cap V_r|}{2}
=
\frac{|N(a)\cap N(b)\cap V_r|(|N(a)\cap N(b)\cap V_r|-1)}{2}.
\end{equation}

%%%%%%

Besides quantity $|N(a)\cap N(b)\cap V_r|$ determines the number of
complete bipartite subgraphs $K_{2,m}$ with vertices $a$ and $b$ and vertices of subset 
\begin{equation}
N_{K_{2,m}}^{(ur)}(a,b)
=
\binom{|N(a)\cap N(b)\cap V_r|}{m}=\frac{|N(a)\cap N(b)\cap V_r|!}
{m!(|N(a)\cap N(b)\cap V_r|-m)!}.
\end{equation}

%%%%%
In the case of $\theta_i=0$, initial state is reduced to
$|0\rangle^{\otimes |V|}$, and the expression for the correlator is as follows
\begin{eqnarray}
\left\langle
\sigma_{a}^{z}\sigma_{b}^{z}
\right\rangle=\frac14\sum_{\beta_a=\pm1}\sum_{\beta_b=\pm1}\prod_{r=1}^{p}\prod_{j\in A_{\mathrm{ext}}\cap V_r}\cos\Phi_{rj}=\nonumber\\
=\frac{1}{2}(\cos\phi)^{|N(a)|+|N(b)|-2|N(a)\cap N(b)|}(1+(\cos2\phi)^{|N(a)\cap N(b)|}).
\label{res}
\end{eqnarray}

Not that the value $|N(a)\cap N(b)|$ is related to  number of
four-cycles $C_4$ with vertices $a$ and $b$ as their opposite
vertices. The number reads
\begin{equation}
N_{C_4}(a,b)
=
\binom{|N(a)\cap N(b)|}{2}
=
\frac{|N(a)\cap N(b)|(|N(a)\cap N(b)|-1)}{2}.
\end{equation}

Thus, quantum correlators are related to structural properties of multipartite graphs, namely the number of
four-cycles.

At the end of this section it is worth noting that in the case of $\theta_i=0$, $\phi_{ij}=\phi$, the mean value
$\left\langle \sigma_a^z \right\rangle$ is related to the degree  of corresponding vertex $a$, $|N(a)|$. It reads
\begin{eqnarray}
\left\langle \sigma_a^z \right\rangle
=
(\cos\phi)^{|N(a)|}.
\end{eqnarray}
So, quantifying $\left\langle \sigma_a^z \right\rangle$, $\left\langle \sigma_b^z \right\rangle$ one can find degrees of vertices $a$, $b$ and then from correlator $\left\langle
\sigma_{a}^{z}\sigma_{b}^{z}
\right\rangle$ we obtain $|N(a)\cap N(b)|$ and the number of four-cicles.

\section{Quantifying stuctural properties of graphs with quantum computing}

Let us quantify the number of closed-four circles with quantum computing. 
For example we consider a quantum state corresponding to  complete $K_{2,3}$ graph.
The state reads

\begin{figure}[H]
\centering
\centerline{\includegraphics[scale=0.65]{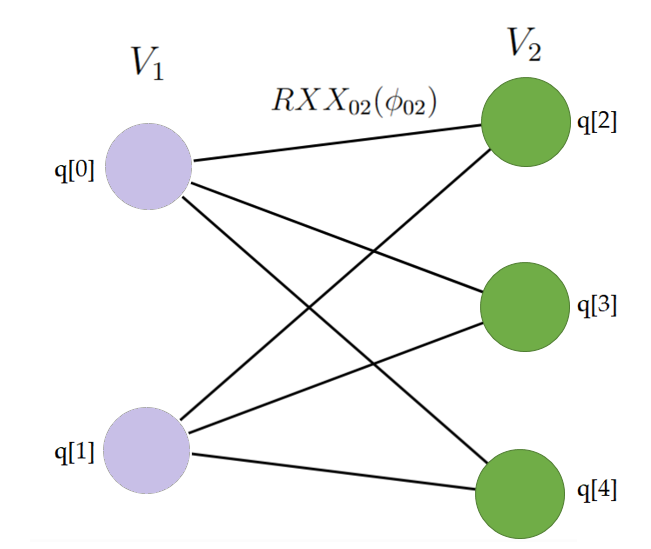}}
\caption{Complete bipartite graph $K_{2,3}$
with 
$V_1=\{q_0,q_1\}$ and
$V_2=\{q_2,q_3,q_4\}$. Each vertex $k$ is represented with qubit $q[k]$, each  eadge (k,l) is represented with action of $RXX_{kl}(\phi_{kl})$ gate on the initial state.}
\label{K23graph}
\end{figure}

\begin{figure}[H]
\centering
\resizebox{\columnwidth}{!}{%
\begin{quantikz}[
row sep=0.35cm,
column sep=0.12cm,
operator/.append style={inner xsep=1pt, inner ysep=2pt}
]
\lstick{$q[0]:\ket{0}$}
& \gate{RY(\theta_0)}
& \gate{RZ(\alpha_1)}
& \gate[3]{RXX_{02}(\phi_{02})}
& \gate[4]{RXX_{03}(\phi_{03})}
& \gate[5]{RXX_{04}(\phi_{04})}
& \qw & \qw & \qw & \qw
\\
\lstick{$q[1]:\ket{0}$}
& \gate{RY(\theta_1)}
& \gate{RZ(\alpha_1)}
& \qw & \qw & \qw
& \gate[2]{RXX_{12}(\phi_{12})}
& \gate[3]{RXX_{13}(\phi_{13})}
& \gate[4]{RXX_{14}(\phi_{14})}
& \qw
\\
\lstick{$q[2]:\ket{0}$}
& \gate{RY(\theta_2)}
& \gate{RZ(\alpha_2)}
& \qw & \qw & \qw
& \qw & \qw & \qw & \qw
\\
\lstick{$q[3]:\ket{0}$}
& \gate{RY(\theta_3)}
& \gate{RZ(\alpha_2)}
& \qw & \qw & \qw
& \qw & \qw & \qw & \qw
\\
\lstick{$q[4]:\ket{0}$}
& \gate{RY(\theta_4)}
& \gate{RZ(\alpha_2)}
& \qw & \qw & \qw
& \qw & \qw & \qw & \qw
\end{quantikz}
}
\caption{Quantum protocol for preparation of quantum state corresponding to the
complete bipartite graph $K_{2,3}$.
Gates
$RY(\theta_i)$, $RZ(\alpha_i)$ are used for preparation of the initial state, 
 action of  $RXX_{ij}(\phi_{ij})$ gates encode
the edges of the graph.}
\label{K23}
\end{figure}
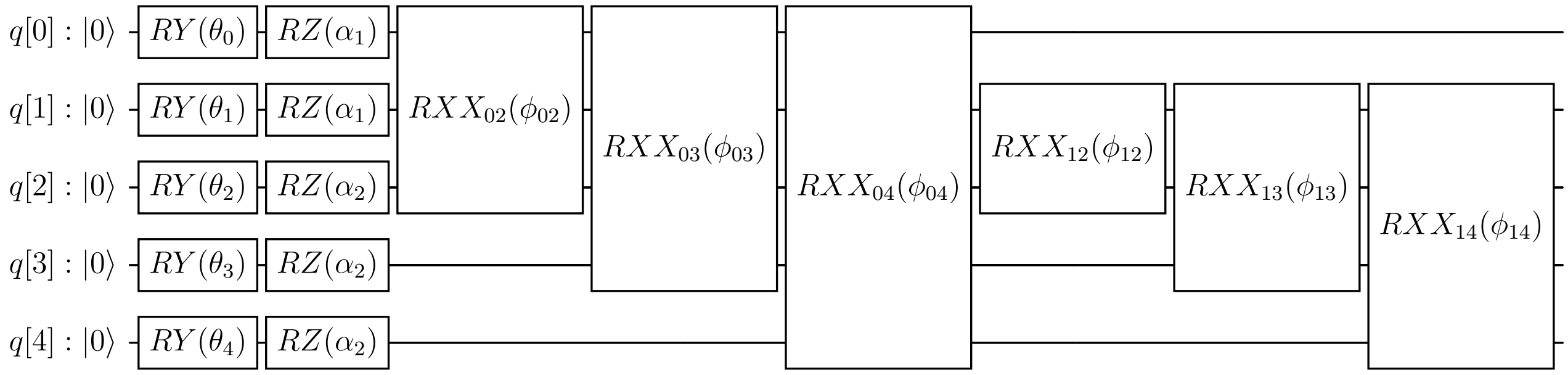

The corresponding five-qubit quantum state reads

\begin{equation}
|\psi_{K_{2,3}}\rangle
=
\prod_{\substack{i\in V_1\\j\in V_2}}
RXX_{ij}(\phi_{ij})
\prod_{r=1}^{2}
\prod_{k\in V_r}
RZ_k(\alpha_r)RY_k(\theta_k)
|00000\rangle,
\end{equation}
where $V_1=\{0,1\}$ and $V_2=\{2,3,4\}$. 
 The state corresponds to $K_{2,3}$ graph presented in Fig.~\ref{K23graph}.
In Fig. \ref{K23} quantum protocol for preparing of the state is presented. 
To quantify the number of four-cycles, let us consider the case
$\theta_k=0$ and $\phi_{ij}=\phi$.
In this case, $\langle\sigma_a^z\rangle
=(\cos\phi)^{|N(a)|}$.
Thus, the degrees of the vertices represented by qubits $q[0]$, 
$q[1]$ can be determined by measuring the single-qubit expectation 
values $\langle\sigma_0^z\rangle$, $\langle\sigma_1^z\rangle$ 
in the state $|\psi_{K_{2,3}}\rangle$. 
The values $|N(0)\cap N(1)|$, $|N(2)\cap N(3)|$ can be quantified 
by measuring the quantum correlators 
$\langle\sigma_0^z\sigma_1^z\rangle$, 
$\langle\sigma_2^z\sigma_3^z\rangle$ and also taking into account (\ref{res}).

For the complete bipartite graph $K_{2,3}$, we have
$|N(0)|=|N(1)|=3$, $|N(2)|=|N(3)|=|N(4)|=2$, 
$|N(0)\cap N(1)|=3$, 
$|N(2)\cap N(3)|=|N(3)\cap N(4)|=|N(2)\cap N(4)|=2$.
We prepared the quantum 
state using the quantum protocol presented in Fig. 1 with $\phi_{ij}=\pi/4$, $\theta_i=0$. 
 The expectation values $\langle\sigma_i^z\rangle$, 
$\langle\sigma_i^z\sigma_j^z\rangle$ were quantified using the AerSimulator, 
which includes a readout error of the order of $10^{-2}$, 
a Pauli-$X$ error of $10^{-4}$, and a CNOT error of $10^{-2}$.
We obtained $\langle\sigma_0^z\rangle=0.305$, 
$\langle\sigma_1^z\rangle=0.320$, 
$\langle\sigma_0^z\sigma_1^z\rangle=0.408$. 
On the basis of these results, we find $|N(0)\cap N(1)|=3.06$. 
Thus, the number of four-cycles $C_4$ with vertices $0$ and $1$ 
as opposite vertices reads $N_{C_4}(0,1)=3$.
We also obtained $\langle\sigma_2^z\rangle=0.369$, 
$\langle\sigma_3^z\rangle=0.457$, 
$\langle\sigma_2^z\sigma_3^z\rangle=0.371$, 
and therefore $|N(2)\cap N(3)|=1.22$. 
 The number of four-cycles $C_4$ with vertices $2$ and $3$ 
as opposite vertices reads $N_{C_4}(2,3)=1$.
The results of quantum computing are in agreement with the 
theoretical results.

 \section{Conclusions}
 We propose a method for constructing quantum states representing edge- and vertex-weighted multipartite graphs. The states are constructed by the action of $RXX$ gates, which determine the edges of the graph, with edge weights related to the parameters of the gates. As initial states, we consider arbitrary separable states of $n$ qubits, each qubit representing a vertex of the graph. These states are prepared by the action of single-qubit rotation gates, whose parameters determine the vertex weights and the membership of the vertices in the corresponding partitions.

 In the general case of quantum states representing multipartite graphs of arbitrary structure, we have calculated quantum correlators and established their relation to the properties of the corresponding graphs. Namely, we have found that quantum correlators in multipartite quantum graph states depend on the distribution of the neighbors of the selected vertices among different partitions of the graph.  The results allow us to relate quantum correlations to important structural properties of multipartite graphs. In particular, on the basis of the obtained results, we propose a method for determining the number of four-cycles containing two selected vertices as opposite vertices. In addition, the results provide a basis for quantifying the number of complete bipartite subgraphs $K_{2,m}$ containing two selected vertices and $m$ vertices belonging to another partition of the graph. Thus, the obtained relations open up possibilities for studying the structural properties of multipartite graphs using quantum computing.

As an example, a quantum state representing the $K_{2,3}$ graph was studied. We quantified the number of four-cycles in the graph, as well as the number of complete bipartite subgraphs, using quantum computing on a noisy IBM quantum simulator. The results of quantum calculations are in agreement with the theoretical ones.

\section{Acknowledgments}

This work was supported by Project 2025.07/0326 (No. 0126U002943) from National Research Foundation
of Ukraine.
ChatGPT-6.2 (OpenAI) was used to improve the English, grammar, and writing style of the manuscript. Author reviewed and approved the final version of the paper and take full responsibility for its content.

\end{document}